\documentclass[12pt]{article}
\usepackage{graphicx}
\usepackage{amsfonts}

\usepackage{epstopdf}

\usepackage{graphics,epsfig}

\usepackage{color}

\def\dalemb#1#2{{\vbox{\hrule height .#2pt
        \hbox{\vrule width.#2pt height#1pt \kern#1pt
                \vrule width.#2pt}
        \hrule height.#2pt}}}

\let\a=\alpha \let\b=\beta \let\g=\gamma \let\d=\delta \let\e=\epsilon
\let\z=\zeta  \let\th=\theta  
\let\l=\lambda \let\m=\mu \let\n=\nu \let\x=\xi  
\let\s=\sigma \let\t=\tau  \let\f=\phi  
 
      \let\G=\Gamma \let\D=\Delta \let\Th=\Theta \let\L=\Lambda
\let\X=\Xi    \let\Y=\Psi
 
\let\la=\label  
  
\def\nn{\nonumber} \def\bd{\begin{document}} \def\ed{\end{document}}
\def\ds{\documentstyle} \let\fr=\frac \let\bl=\bigl \let\br=\bigr
\let\Br=\Bigr \let\Bl=\Bigl
\let\bm=\bibitem
\let\na=\nabla
\def\tU{{\widetilde U}}
\let\pa=\partial \let\ov=\overline
\def\ie{{\it i.e.\ }}
\newcommand{\be}{\begin{equation}}
\newcommand{\ee}{\end{equation}}
\def\ba{\begin{array}}
\def\ea{\end{array}}
\def\ft#1#2{{\textstyle{{\scriptstyle #1}\over {\scriptstyle #2}}}}
\def\fft#1#2{{#1 \over #2}}
\def\F#1#2{{ F_{#1}^{(#2)} }}
\def\cF#1#2{{ {\cal F}_{#1}^{(#2)} }}

\def\R{{\bf R}}
\def\sst#1{{\scriptscriptstyle #1}}
\def\oneone{\rlap 1\mkern4mu{\rm l}}
\def\e7{E_{7(+7)}}
\def\td{\tilde}
\def\wtd{\widetilde}
\def\im{{\rm i}}
\def\bog{Bogomol'nyi\ }
\newcommand{\ho}[1]{$\, ^{#1}$}
\newcommand{\hoch}[1]{$\, ^{#1}$}
\newcommand{\bea}{\begin{eqnarray}}
\newcommand{\eea}{\end{eqnarray}}
\newcommand{\ra}{\rightarrow}
\newcommand{\lra}{\longrightarrow}
\newcommand{\Lra}{\Leftrightarrow}
\newcommand{\ap}{\alpha^\prime}
\newcommand{\bp}{\tilde \beta^\prime}
\newcommand{\cB}{{\cal B}}
\newcommand{\cO}{{\cal O}}
\newcommand{\vecx}{\vec{x}}
\newcommand{\vecy}{\vec{y}}
\newcommand{\vecp}{\vec{p}}
\newcommand{\vecq}{\vec{q}}
\newcommand{\tr}{{\rm tr} }
\newcommand{\Tr}{{\rm Tr} }
\newcommand{\NP}{Nucl. Phys. }

\newcommand{\cL}{{\cal L}}
\newcommand{\cA}{{\cal A}}

\newcommand{\cD}{{\cal D}}

\def\Ht{\tilde{H}}

\def\sst#1{{\scriptscriptstyle #1}}
\def\0{{\sst{(0)}}}
\def\1{{\sst{(1)}}}
\def\2{{\sst{(2)}}}
\def\3{{\sst{(3)}}}
\def\4{{\sst{(4)}}}
\def\5{{\sst{(5)}}}
\def\6{{\sst{(6)}}}
\def\7{{\sst{(7)}}}
\def\8{{\sst{(8)}}}
\def\9{{\sst{(9)}}}
\def\ve{\varepsilon}
\def\vf{\varphi}
\def\F{\Phi}
\def\wg{\wedge}

\newcommand{\tamphys}{\it 
}

\newcommand{\auth}{AUTHORS}

\def\Lab{\bar{\L}}
\def\Psib{\bar{\Psi}}
\def\Delb{\bar{\D}}
\def\Thb{\bar{\Theta}}
\def\Sigb{\bar{\Sigma}}

\def\sib{\bar{\s}}

\def\mub{\bar{\mu}}
\def\ztb{\bar{\zeta}}
\def\psib{\bar{\psi}}
\def\thb{\bar{\theta}}
\def\chib{\bar{\chi}}

\def\ab{{\bar{a}}}
\def\Ab{\bar{A}}
\def\pb{\bar{p}}
\def\qb{\bar{q}}
\def\cb{\bar{c}}
\def\db{\bar{d}}
\def\zb{\bar{z}}
\def\Zb{\bar{Z}}

\def\wb{\bar{w}}
\def\lb{\bar{\l}}
\def\Jb{\bar{J}}
\def\Nb{\bar{N}}
\def\pab{\bar{\pa}}
\def\etab{\bar{\eta}}
\def\phib{\bar{\phi}}

\def\e{\epsilon}

\def \bi{\bibitem}
\def \la {\label}

\def \l {\lambda}
\def\foot{\footnote}
\def \tl  {{\tilde \l}}
\def \sql {{\sqrt \l}}
\def \adss {$AdS_5 \times S^5$\ }
\newcommand{\rf}[1]{(\ref{#1})}
\def \ov {\over}

\def\th{\theta}
\def\Th{\Theta}
\def\vth{\vartheta}
\def\btheta{{\bar\theta}}
\def\ttheta{{{\tilde\theta}}}
\def\bttheta{{{\bar\ttheta}}}
\def\vth{\vartheta}

\def\ra{\rightarrow}

\def\au{{\underline{a}}}
\def\bu{{\underline{b}}}
\def\ual{\underline{\a}}
\def\ube{\underline{\b}}
\def\uM{\underline{M}}
\def\uN{\underline{N}}
\def\uP{\underline{P}}
\def\muu{\underline{\mu}}

\def\cc{\circ}
\def\eqv{\equiv}

\def\ni{\noindent}

\def\Ep{E^{{}^{(+)}}}
\def\Em{E^{{}^{(-)}}}

\def\Mp{M^{{}^{(+)}}}
\def\Mm{M^{{}^{(-)}}}

\def \ha{{1\ov 2}}

\def\r{\rho}

\def\Y{{\rm Y}}
\def\X{{\rm X}}
\def\tY{\tilde{\rm Y}}
\def\tX{\tilde{\rm X}}
\def\dY{\dot{\rm Y}}
\def\dX{\dot{\rm X}}

\def \J {\mathcal{J}}
\def \del {\partial}

\def\dF{\dot{F}}
\def\dG{\dot{G}}
\def\adot{\dot{a}}
\def\bdot{\dot{b}}
\def\df{\dot{f}}

\def\dal{{\dot{\alpha}}}
\def\dbe{{\dot{\beta}}}
\def\dga{{\dot{\gamma}}}

\def \E {{\cal E}}
\def \cS {{\cal S}}
\def \J {{\cal J}}
\def\N{{\cal N}}
\def\F{{\cal F}}
\def\M{{\cal M}}

\def\ms{\mathcal{S}}
\def\mj{\mathcal{J}}
\def\soj{\fr{\ms}{\mj}}
\def \R {{\bf R}}
\def \om {\omega}
\def \bE {\bar E}
\def \x {{\cal X}}

\def \bi{\bibitem}
\def \la {\label}

\def \l {\lambda}
\def\foot{\footnote}
\def \tl  {{\tilde \l}}
\def \sql {{\sqrt \l}}
\def \adss {$AdS_5 \times S^5$\ }
\def \ov {\over}

\def \varpi {{\rm w}}

\def\e{\epsilon}

\def\At{\tilde{A}}
\def\Bt{\tilde{B}}
\def\Ct{\tilde{C}}
\def\Dt{\tilde{D}}
\def\Et{\tilde{E}}
\def\Ft{\tilde{F}}
\def\Gt{\tilde{G}}
\def\Mt{\tilde{M}}
\def\at{{\tilde{a}}}
\def\bt{{\tilde{b}}}
\def\ct{\tilde{c}}
\def\dt{\tilde{d}}
\def\et{\tilde{e}}
\def\ft{\tilde{f}}
\def\gt{\tilde{g}}
\def\chit{\tilde{\chi}}
\def\psibt{\tilde{\psib}}
\def\mut{\tilde{\mu}}
\def\mubt{\tilde{\mub}}
\def\wt{\tilde{w}}
\def\wbt{\tilde{\wb}}

\def\ola{\overleftarrow}
\def\ora{\overrightarrow}
\def\alt{\tilde{\a}}

\def\dh{\hat{d}}
\def\bh{\hat{b}}
\def\deh{\hat{\d}}
\def\mh{\hat{m}}
\def\nh{\hat{n}}

\def\ah{\hat{a}}
\def\eh{\hat{e}}
\def\Eh{\hat{E}}
\def\eph{\hat{\e}}
\def\ph{\hat{p}}
\def\Ah{\hat{A}}
\def\alh{\hat{\a}}
\def\beh{\hat{\b}}
\def\gah{\hat{\g}}
\def\muh{{\hat{\m}}}
\def\nuh{{\hat{\n}}}
\def\thh{\hat{\th}}
\def\dh{\hat{d}}
\def\ih{\hat{i}}
\def\jh{\hat{j}}
\def\deh{\hat{\d}}
\def\uh{\hat{u}}
\def\vh{\hat{v}}
\def\wh{\hat{w}}
\def\lah{\hat{\l}}
\def\Ch{\hat{C}}
\def\Omh{\hat{\Omega}}
\def\rh{\hat{r}}
\def\sh{\hat{s}}
\def\that{\hat{t}}
\def\lah{\hat{\l}}

\def\dgg{\dagger}

\def\ps{\rlap{\, /}\;\,p }
\def\ks{\rlap{\, /}\;\,k }
\def\Ds{\rlap{\, /}\;\,D }
\def\Hs{\rlap{\, /}\;H }

\def\gym{g_{YM}}

\def\bpa{\bar{\pa}}

\newcommand{\bes}{\begin{subequations}}
\newcommand{\ees}{\end{subequations}}

\newcommand{\eda}{\bar{\eta}}
\newcommand{\teta}{\tilde{\eta}}
\newcommand{\tta}{\tilde{\eta}}
\newcommand{\sP}{/\!\!\!\partial}
\newcommand{\sF}{\: /\!\!\!\! F}

\newcommand{\dd}{{d}}

\newcommand{\dsfrac}{\displaystyle\frac}

\begin{document}
\overfullrule=0pt
\parskip=2pt
\parindent=12pt
\headheight=0in \headsep=0in \topmargin=0in
\oddsidemargin=0in

\vspace{ -3cm}
\thispagestyle{empty}



 \vspace{0.1cm}

\setcounter{equation}{0}
\setcounter{footnote}{0}
\setcounter{section}{0}



\begin{center}

{\Large\bf $N^3$ entropy of M5 branes from dielectric effect
  }

\vskip 0.8cm

 \vspace{.5cm}

E. Hatefi\\

{ {
{\it International Centre for Theoretical Physics\\
Strada Costiera 11, Trieste, Italy. \\
ehatefi@ictp.it
}}}

\vspace{0.6cm}
A. J. Nurmagambetov\\
{\it A.I. Akhiezer Institute for Theoretical Physics of NSC KIPT\\
1 Akademicheskaya St., Kharkov, UA 61108, Ukraine\\
{
ajn@kipt.kharkov.ua}
}


\vspace{0.5cm}
I. Y. Park
\\



{\it Department of Natural and physical Sciences,
Philander Smith College 
                               \\
Little Rock, AR 72223, USA \\
inyongpark05@gmail.com
}

\end{center}

 \vspace{0.1cm}

 \begin{abstract}
We observe that the $N^3$ entropy behavior of
near-extermal M5 branes
can be reproduced from SYM side with the role of Myers' terms. We start by generalizing the
Klebanov-Tseytlin (KT) supergravity
solution that displays the $N^3$ entropy behavior. The new feature of the general solution is visibility of
the "internal" degrees of the M5 branes, i.e., the M0 branes and
the M2 branes. With the rationale provided by the supergravity
analysis, we consider a D0 brane quantum mechanical setup with Myers'
terms. Using localization technique, we show that the leading
$N^3$ behavior of the free energy comes from the "classical contribution"
with the rest sub-leading.

\end{abstract}
\newpage

\setcounter{equation}{0}
\setcounter{footnote}{0}
\setcounter{section}{0}


\section{Introduction}

One of the recent lessons on the mechanism behind AdS/CFT - that was
born in the birthplace of D-brane physics \cite{Polchinski:1995mt} - is that
it may be far more dynamical in nature than
might have originally been conceived. Although the complete first-principle derivation of the correspondence
is still missing, indications exist that such dynamical nature
would be the key to the derivation. One aspect of the dynamical nature
is (conjectured) generation of D-brane curvature out of the non-geometric
theories (SYM/open string) that one starts with. It was
proposed in \cite{Park:2007mc} that open string {\em quantum effects}
may engineer the curvature around the host branes.
Transition between an open
string and a closed string is another dynamical aspect, which
should also play a crucial role \cite{Park:2001bm} in the first-principle derivation of AdS/CFT.

 In the original formulation
of AdS/CFT, it was SYM, the massless modes of the open
string, that received attention after a "decoupling"
limit. It has been increasingly clear to our view, however, that it is the {\em full-fledged}
open string theory that is required for reproduction of the results of
the dual closed string in general. Although this would be at odds with
the strongest (or most radical) form of the AdS/CFT, it should not be
entirely surprising since the correspondence would then be understood
as a generalization of open-closed string type duality \cite{Park:2001bm}. In one further step,
we point out a possibility below which would take the open string
to a much elevated status. This possibility is that the role of the open string is more
fundamental than might have been expected: {\em an open string} setup may be
required to reproduce certain
 {\em near-extremal supergravity} results.\footnote{An open string
ideal gas model
was used to reproduce the $N^2$ entropy behavior of a near-extremal
D3 brane configuration \cite{Gubser:1996de}. In the present case, it is
the closed string interaction terms, i.e., the Myers' terms, that are essential
for reproduction of $N^3$ entropy. }

A large amount of
evidence has been accumulated over the years, especially in the case of AdS$_5$/CFT$_4$.  Nevertheless, we note in this paper that there could be bulk physics associated with stringy effects that may not be suitably described by pure SYM even after non-perturbative effects
are taken into account. ("Pure" SYM means SYM without an extra effect such as the dielectric effect.) Non-extremality of certain supergravity
solutions may reflect massive open stringy effect \cite{Klebanov}\cite{Park:1999xz},
 and potentially be an example. More specifically,
the entropy of such a configuration will reflect stringy effects, and may take more than pure SYM on the dual side. Naturally one may wonder whether/how
such effects can be seen from the SYM side.

In this paper, we take the entropy of various near-extremal M5
brane configurations; there have been suggestions that 5D
maximal SYM may adequately describe the M5 brane dynamics \cite{Douglas:2010iu}\cite{Lambert:2010iw}. The $N^3$ behavior was observed
in \cite{Henningson:1998gx} in the field theory anomaly context.
The authors of
\cite{Kim:2011mv} considered the quantum mechanical
model that results from reducing the 5D SYM to time dimension.
(See \cite{Lee:2000hp}\cite{Bolognesi:2011rq}\cite{Bolognesi:2011nh} for related discussions.)
They studied the model's degrees of freedom, and computed a certain
index of the model.
5D SYM theory is the low energy limit of D4 branes, and, as such, a
UV completion is required for a proper description of the D4 dynamics. For example,
it potentially has an issue with renormalizability. In this work, we largely
set this issue aside because we will take an alternative route.

It is natural to believe that the Kaluza-Klein modes (see, e.g., \cite{Kim:2011mv}) and the self-dual string (see, e.g., \cite{Howe:1997ue}\cite{Berman:2007bv})
should be responsible for the $N^3$ behavior of M5 branes. (They are "instanton particles"
and "little" strings in the IIA
setup respectively.) We propose below using the IIA setup that
  it is the dielectric effect \cite{Emparan:1997rt}\cite{Myers:1999ps} in a D0/D4 system that is responsible for
the leading $N^3$ entropy behavior. A D0/D4 system can be studied by the quantum mechanical lagrangian that results from reducing the $\N=1$
6D gauge theory whose field contents are vector multiplet,
adjoint hypermultiplet and fundamental hypermultiplet, to time dimension.
The Myers' terms enter in the standard manner. The vacuum structure
was studied in the Myers' paper \cite{Myers:1999ps};
we will show that the non-commuting solution
is responsible for the $N^3$ behavior.

To establish the entropy correspondence between supergravity and
open string/SYM, it is necessary to understand the mapping
between various branches of moduli space in these theories: physics of a
given {\em branch} of supergravity moduli space must be matched
with physics of the corresponding {\em branch} of SYM.
  In other words, AdS/CFT mappings of two theories
should be made {\em branch by branch}.
Once the dielectric effect is taken into account, the minimum
energy configuration is a non-commuting solution.
One must consider the non-commutative
branch (as opposed to, e.g., the branch that is associated with a commuting
solution) because that would correspond to the supergravity moduli
branch under consideration here.\footnote{
The whole issue is tied with the limitations of our tools (both in open string/SYM and
supergravity) that do not, in general, allow one to choose the degrees of freedom
appropriate for different branches "in a continuous manner".
}

An important point to be made in the following sections is
how to justify the use of the one dimensional SYM.
The 5D SYM has generalized instanton-type solutions with
various moduli. It will be argued that as a result of the
branch mapping between SYM and supergravity, one must consider point-like
soliton solutions.
With that argument, we will consider the reduced action, i.e., D0 action, since
the D0 brane would correspond to the point-like "instanton" solutions\footnote{
See \cite{Dorey:2002ik} and \cite{Vandoren:2008xg} for reviews of instanton.
},
a narrower than otherwise branch of the SYM moduli space.

One of the interesting aspects of D-brane physics is that
it admits various "dual" descriptions. Let us illustrate
this "duality" taking the present example, a D0/D4 system.
There are two ways to describe the system within the field theory
technique, the D4-based description and the D0-based
description. (Interesting discussions
can be found in \cite{Hashimoto:2005qh} and \cite{Ito:2009ac} on related matters.)
 In the D4-based description, one uses 5D SYM, and D0 branes are
realized as its soliton solutions, the "instantonic particles".
If the 5D SYM theory were complete, one would integrate out the perturbative degrees of freedom,
and get the D0 action at an intermediate stage of computing the partition function.
One would then evaluate the resulting action further to complete the partition function.
However, since 5D SYM is not (known to be) complete (see \cite{Young:2011aa} for
recent progress),
one should rely on the full open string techniques. In the worldsheet
setup of the open string, it is not entirely clear how to integrate out the perturbative brane degrees of freedom. (One way would be to use field theory techniques including all the massive modes of the open string. However, this procedure would be impossible to implement.)
Therefore, technical reasons alone force one to turn to the fundamental description of D0
branes, D0-based description of D0/D4 system (more on this later).

One thing that needs to be understood in the D0 based
description is how D0/D4 system is different from pure D0 system.
In other words, the blow-up type solution is there even for a purely
D0 brane system. What differentiates D0/D4 system from
pure D0 system in the D0 based description?
The answer to this question may lie in the peculiar string theoretic manner that the dielectric effect
enters as we will point out later.

The near-extremal limit of supergravity
should correspond to keeping the next leading order terms in $\a'$. Therefore, one should
take the corresponding step in the
SYM side; in general, it is expected that Myers' terms will
be an infinite series in a derivative expansion.
Taking the near-extremal limit in the supergravity configuration should correspond to taking the leading terms out of the infinitely many
 Myers' terms.

\vspace{.3in}

The organization of the rest of the paper is as follows. In section 2, after briefly reviewing the
Klebanov-Tseytlin (KT) solution \cite{Klebanov:1996un}, we construct a class of IIA (or 11D)
supergravity solutions of a D0/D4 type that display $N^3$ behavior by putting together ingredients
in literature. The solutions can be viewed as a generalization of the Klebanov-Tseytlin solution.
We observe that D2 branes are present through the Myers type effect. We take the existence of such solutions as an indication that the Myers' effect should be included in the gauge theory description as well.
 In section 3, we set the stage for section 4
 in which we analyze the entropy
of the D0/D2/D4 system in a D0 brane-based description.
We determine in our convention the 5D SYM lagrangian with
the Myers term, a system that yields a quantum mechanical model
with the D4 brane effect and Myers' effect incorporated.
The action is reduced to time dimension.
With the quantum mechanical lagrangian
ready, we compute the entropy of the system through localization
technique \cite{Witten:1988ze}\cite{Kapustin:2009kz}. To that end, we first recall facts about the residual supersymmetry
of the system. After noting that the partition function
consists of several parts,
it is shown that the $N^3$ behavior comes from the
"classical part"\footnote{The Myers terms come from open strings
coupling to a closed string. Therefore, it can effectively be view as
a loop effect from the standpoint of the open strings.} with
the other parts yielding sub-leading contributions.
In section 5, we discuss some of the subtle issues, and comment
on future directions.

\section{Rationale provided by supergravity}

Although the KK modes and the self-dual strings are likely to be
responsible for the $N^3$ behavior, their presence
in the near extremal solutions \cite{Gueven:1992hh}\cite{Duff:1996hp}\cite{Klebanov:1996un}
is not evident.
 There might exist a more general class of solutions
with the same $N^3$ behavior. It is expected that they would have
the following characteristics: reduction to the KT solution in some limit,
and the more evident presence of the KK modes and the self-dual strings (or M2 branes).

In this section, we confirm this expectation by explicit
construction of a class of near-extremal solutions
of D0/D4 with the $N^3$ entropy behavior.
They indeed reduce to the KT solution in a certain limit, and provide rationale
for incorporating the Myers' term in the open string/SYM analysis in the following
sections. Presumably, the near extremal M5 solution of \cite{Klebanov:1996un}
could be viewed as describing M5 branes with the lower dimensional M-branes completely
dissolved.

\vspace{.3in}

One can construct a solution that describes a KKW/M2/M5 by boosting an M2/M5 solution \cite{Izquierdo:1995ms}\cite{Russo:1996if}\cite{Costa:1996re}
along, say, the $x_3$ direction. (Details are presented in \cite{NP}.) The resulting solution has the following form
\[
ds^2_{11}=(H\Ht)^{1/3}\left[H^{-1}(-K^{-1}fdt^2+d{ x}^2_1+dx^2_2)+\Ht^{-1}(Kd{\hat x}^2_3+dx^2_4+dx^2_5) \right.
\]
\be
\left. +f^{-1}dr^ 2+r^2d \Omega^2_4\right] ,
\la{WM2M5int}
\ee
where
\[
H=1+N\fr{h_0^3}{r^3},\quad \Ht=\sin^2 \zeta+H\cos^2 \zeta,\quad  f=1-\fr{\m^3}{r^3},
\]
\be
K=1+N_0 \fr{k_0^3}{r^3},\quad  d{\hat x}_3^2=[dx_3+(K^{-1}-1)dt]^2 ,
\la{WHHtfint}
\ee
{and
\be
\hat{F}_4=\fr12\cos\zeta \ast dH+\fr12\sin \zeta \,dH^{-1}\e_3+\fr32\sin2\zeta \,H^{-2} dH \, \bar{\e}_3.
\ee
$\e_3$ and $\bar{\e}_3$ are volume forms on $\mathbb{M}^3$ and $\mathbb{E}^3$ parameterized respectively by $(t,x_1,x_2)$ and $(x_3,x_4,x_5)$; $\ast$ is the Hodge dual of $\mathbb{E}^5$ that is transverse to the M5 branes.}

This solution is a generalization of the boosted solution of \cite{Russo:1996if}\cite{Costa:1996re}.
In $N h_0^3 \ll 1$, $N_0 k_0^3 \gg 1$ limit, the solution \rf{WM2M5int} can be viewed as a stack of $N_0$ black M5-branes. This should be a manifestation of the Myers' effect, since,
in the supergravity context, the dielectric effect should manifest itself
as a dissolution of lower dimensional
branes into higher dimensional ones.
One can show that the solution exhibits the $S\sim N^3 T^5$ entropy behavior {in the near-extremal limit}.

Since we will use IIA setup in the following sections, let us reduce the solution to the corresponding
IIA configuration. The resulting configuration will have the same {near-extremal} entropy behavior { $S\sim N^3 T^5$}.
Dimensional reduction of \rf{WM2M5int} along {the $x_3$ direction in which KK waves travel}
 leads to the D0/D2/D4 interpolating solution, D2/D4 part of which was constructed in \cite{Green:1996vh} in the string frame.
{In the Einstein frame,
\bea
ds^2_{11}=e^{-\phi/6}ds^2_{10}+e^{4\phi/3}(dx_3+A)^2,
\eea
reducing along $x_3$, one gets
\[
ds^2_{10}=H^{3/8}\Ht^{1/4}K^{-3/8}\left[H^{-1}(-K^{-1}fdt^2+dx_1^2+dx^2_2)+\Ht^{-1}(dx^2_4+dx^2_5) \right.
\]
\be
\left. +f^{-1}dr^2+r^2d\Omega^2_4 \right] ,
\la{D0D2D4int}
\ee
\be
e^\phi=H^{1/4}\Ht^{1/2}K^{-3/4},\qquad A_{[1]}=(K^{-1}-1)dt ,
\la{D0D2D4phi}
\ee
and, on account of $\hat{F}_4=F_4+F_3 \wg  (dx_3+A)$,
\be
F_4=\fr12\cos\zeta \ast dH+\fr12\sin \zeta \,dH^{-1}\e_3,\quad F_3=\fr32\sin2\zeta \,H^{-2} dH \, \bar{\e}_2.
\la{D0D2D4F}
\ee
$\e_3$ and $\bar{\e}_2$ are the $(t,x_1,x_2)$ and $(x_3,x_4)$ coordinate volume forms respectively; $\ast$ still the Hodge dual in the transversal $\mathbb{E}^5$.} Setting $K=1$ {in  \rf{D0D2D4int}--\rf{D0D2D4F}} leads to a D2/D4 interpolating solution that generalizes the solution found in
 \cite{Green:1996vh}.

\section{Open string/SYM setup}

Above, we have considered the entropy of the Klebanov-Tseytlin solution and its generalization.
In this section, we review open string derivation of the corresponding
low energy SYM action, focusing on the standard cubic Myers' terms.
The Myers' terms will play a central role in reproducing the $N^3$ entropy
from the SYM side, and the precise form of the SYM action with the Myers' term will be given within our convention.
The action will preserve part of the supersymmetry. The residual supersymmetry will be used in the next section in which the partition function is evaluated using localization technique, a very convenient tool for evaluating the full partition function.

In the D0-based description, the presence of D4 branes are realized in part through the sector that comes from dimensional reduction of the $\N=1$ 6D SYM fundamental
hypermultiplet. (We will have more on this in section 4.) The RR gauge field $C^\3$ enters as a background in the effective
field theory level. One of the key issues is the residual supersymmetry
of the Myers' term since residual susy is essential for
employing localization technique.
In the following subsection, we recall a few things about the residual supersymmetry of
the 5D SYM with the Myers' terms added, and reduce the system to time dimension. With these
tasks completed, we will be ready to compute the partition function
in section 4.

\subsection{Myers' terms, supersymmetry and reduction}

It is convenient to start in six dimensions, and reduce the system to five
dimensions.
The 6D theory has (1,1) supersymmetry. In terms of (0,1) supersymmetry,
the (1,1) vector multiplet splits into an (0,1) vector multiplet, $(A_{\muh},\chi)$,
and a (0,1) hypermultiplet in the adjoint, $(\z,Z)$.
In addition, the system contains the fundamental
 hypermultiplet, $(w,\m)$ \cite{Douglas:1996uz}.

The precise form
of the low energy 5D SYM action that reproduces the corresponding open string results can be determined
by going through the standard procedure. In our convention, it is
given by
\bea
\int \cL &=&\int d^5x\; \Tr\Big[
     -\fr14 F_{\muh\nuh}F^{\muh\nuh}+ \fr12\sum_{i=1}^3\cD_{i}^2
  -{i}\chib_{\dal}\G^{\muh}D_{\muh}\chi^{\dal}
 -(D_{\muh}Z^{\dal})^\dgg  D_{\muh}Z^{\dal}
            -i\ztb \G^{\muh}D_{\muh}\z  \nn\\
 &&  +\Zb_{\dal} \sum_{c=1}^3( \t^c\cD_{c})^{\dal}{}_{\dbe} Z^{\dbe}    -Z_{\dal} \sum_{i=1}^3 ( \t^i\cD_{i} )^{\dal}{}_{\dbe} \Zb^{\dbe}
 +4i\, ([\ztb,Z^{\dbe}]\e_{\dal \dbe}\chi^{\dal}
       -\chib_{\dal}[\z,\Zb_{\dbe}]\e^{\dal \dbe})\;\;
   \Big] \nn\\
&&+ \e^{\dal\dbe}(D_{\muh}\wb_{\dal})_a (D^{\muh}w_{\dbe})^a
       -\wb_{\dal a} \sum_{i=1}^3 (\t^i)^{\dal}{}_{\dbe }
              (\cD_{i})^a{}_b w^{\dbe b} -{i}\mub_a \G^{\muh} (D_{\muh} \m)^a \nn\\
  && +{{2}}i\, [\mub_a(\chi^{\dal})^a{}_b \e_{\dal \dbe} (w^{\dbe})^b
       + (\wb_{\dal})_a \e^{{\dal}\dbe}(\chib_{\dbe})^a{}_b \,\m^b]
  \label{sym3q}
\eea
where
\bea
 \muh=(\m,5)
\eea
For technical reasons, the action is written in the 6D notation.
Other than minor conventional differences, this action can be easily deduced
from the results that were obtained in \cite{DiVecchia:2011mf} which was based on
earlier work of \cite{Billo:2002hm}.
Again, for technical and
convenience reasons, we start with
the 5D SYM above, and reduce it to time dimension
instead of directly considering a D0 system from the beginning.

One feature of the susy transformation of
\rf{sym3q}\footnote{The supersymmetry transformations can be found in, e.g.,
\cite{Dorey:2002ik} and \cite{DiVecchia:2011mf}.} is worth noting: the gauge multiplet fields transform within themselves
(whereas the hypermultiplet transformations involve the
gauge fields). This feature of the susy transformations will be used in the next section where localization is employed to evaluate the partition function.

Let us (implicitly) reduce the action \rf{sym3q} to time dimension, and add the Myers terms\footnote{For supersymmetry in the presence of the Myers terms, see (\cite{Moore:1998et} and) \cite{Austing:2003kd}.}
\bea
  - \fr{i}3  f\,\gym\,\e_{pqr}\f^p\f^q\f^r \label{M}
\eea
 where $f$ is a constant. For now, we will be implicit about the ranges of the indices
  $(p,q,r)$. They will become clear in section 4.2. The resulting action is
\bea
\int \cL_{D0/D4} &=&\int dt\; \Tr\Big[
     -\fr14 F_{\muh\nuh}F^{\muh\nuh}+ \fr12\sum_{i=1}^3\cD_{i}^2
  -{i}\chib_{\dal}\G^{\muh}D_{\muh}\chi^{\dal}
 -(D_{\muh}Z^{\dal})^\dgg  D_{\muh}Z^{\dal}
            -i\ztb \G^{\muh}D_{\muh}\z  \nn\\
 &&  +\Zb_{\dal} \sum_{c=1}^3( \t^c\cD_{c})^{\dal}{}_{\dbe} Z^{\dbe}    -Z_{\dal} \sum_{i=1}^3 ( \t^i\cD_{i} )^{\dal}{}_{\dbe} \Zb^{\dbe}
 +4i\, ([\ztb,Z^{\dbe}]\e_{\dal \dbe}\chi^{\dal}
       -\chib_{\dal}[\z,\Zb_{\dbe}]\e^{\dal \dbe})\nn\\
 && \hspace{2in}
  - \fr23 i f\,\gym\,\e^{pqr}\f_p\f_q\f_r   \Big] \nn\\
&&+ \e^{\dal\dbe}(D_{\muh}\wb_{\dal})_a (D^{\muh}w_{\dbe})^a
       -\wb_{\dal a} \sum_{i=1}^3 (\t^i)^{\dal}{}_{\dbe }
              (\cD_{i})^a{}_b w^{\dbe b} -{i}\mub_a \G^{\muh} (D_{\muh} \m)^a \nn\\
  && +{{2}}i\, [\mub_a(\chi^{\dal})^a{}_b \e_{\dal \dbe} (w^{\dbe})^b
       + (\wb_{\dal})_a \e^{{\dal}\dbe}(\chib_{\dbe})^a{}_b \,\m^b]
\label{sym4}
\eea
We record the potential part of the fields in the adjoint representation
 for later use:
\bea
 &&- \Tr\Big(
\fr14[\f_{\muh},\f_{\nuh}]^2+ \fr12\sum_{c=1}^3\cD_{c}^2
   +\fr12[\f_{\muh},\f_m]^2
+\fr{i}2  \sum_{c=1}^3 \cD_{c}\eta_{cmn} [\f^m,\f^n]
\nn\\
 && \quad\quad  \hspace{2.5in}
  - \fr{i}3  f\,\gym\,\e_{pqr}\f^p\f^q\f^r   \Big)
\label{symd0fromsym3}
\eea
where we have rewritten the $Z$-fields in terms of $\f$'s \cite{DiVecchia:2011mf}.
$\eta_{cmn}$ is the `t Hooft symbol whose explicit form can be found, e.g., in \cite{Billo:2002hm}.
The index $m$ (with $\muh=5$) represents the directions transverse to the D4 branes.
Note that the coefficient of the Myers' terms have
an extra power of the gauge coupling. This is because
they come from the coupling between open string states and
a closed string state. As well-known, the closed string coupling
goes $g_c\sim g_0^2\sim \gym^2$. The extra factor of $\gym$
will be important for the entropy computation later.

\section{Entropy analysis with Myers' term }

With the stage set in the previous sections, we evaluate here
the free energy of the quantum mechanical system, \rf{sym4}.
There are several essential ingredients in the analysis; strictly
speaking, some of them are assumptions.
First of all, the D0-based setup itself is assumed
to properly describe the D0/D4 system with the dielectric effect taken into account. The second ingredient is the aforementioned
issue of the {\em moduli branch matching}
between SYM and supergravity. In the present context, we take this to
imply that the SYM branch corresponding to the supergravity branch
under consideration should be associated with $w_0=0$, where $w_0$ denotes
the vev of $w$.\footnote{
In principle, the path integral over the quantum fluctuations
of the fundamental hypermultiplet must be considered. However,
 localization technique renders this step unnecessary, as we will see shortly.
}
The reason is that the KT supergravity solution has
the lower dimensional objects completely dissolved: it should correspond to
a situation where the solitons become point-like.
The third ingredient is localization technique, and it is the main topic
of this section: we employ localization technique to evaluate the partition function, and thereby derive the $N^3$ behavior.
Towards the end of this section, we comment on the difference
between a D0/D4 and a D(-1)/D3 system, pondering the peculiar
way in which the Myers' terms arise for Dp branes with $p\neq 0$.
An additional effect of D4 branes in the D0-based
description will be commented on at the end of this section as well.

\subsection{Entropy of D0/D4 via localization}

As recalled in the previous section, the dielectric effect preserves
part of the supersymmetry. With the supersymmetry partially preserved,
one can rely on localization technique\footnote{
In \cite{Austing:2003kd}, it was shown that the supercharge they considered
is nilpotent on gauge invariant quantities. Presumably, once one
adds the gauge fixing term it would make (as in \cite{Kapustin:2009kz}) the sum of the supercharge and
the BRST charge nilpotent
on all quantities, not only on gauge invariant quantities.} to compute
the free energy.
More specifically, consider adding a localizing term $QS_L$ to the action
\rf{sym4},
\bea
 \int D\Phi\; \exp\Big({\fr{i}{\gym^2}S+i\fr{t}{\gym^2}QS_L} \Big)
     \label{taction}
=\exp\Big({\fr{i}{\gym^2}}{\cal F}\Big)
\eea
where $\Phi$ is a collective symbol for the fields, $t$ is a localization parameter and Q is a nilpotent
operator that is constructed basically out of the residual supersymmetry
transformations. The functional $S_L$ represents an appropriately chosen
localizing term; it is possible \cite{Moore:1998et}\cite{Austing:2003kd} to choose
$S_L$ such that the $QS_L$ becomes the action for the adjoint
fields (\rf{localL} below).

In the path integral\footnote{The supergravity solutions found in sec 2 are black brane solutions with finite Hawking temperature. This suggests, although not without subtlety, a possibility that it may be a finite temperature field theory that needs to be employed. We expect the finite temperature to preserve the $N^3$ behavior. The constant $f$ that appears in \rf{M} may be related to the temperature. We postpone the precise relevance and effects of finite temperature for near-future research. }, there are two contributions: the scalar vevs and the one-loop contributions.
The scalar vevs are determined by minimizing the potential with the
Myers' terms. In general, there could be some parameters over which
a matrix theory-type path integral would have to be further
performed. We will note that there is no modulus in the minimum
solution of \cite{Myers:1999ps}, therefore, this step is not required.
As usual, the action will be expanded around the scalar vevs. The one-loop contribution
can be evaluated by the saddle point method.

 The fields in the adjoint representation can be re-expressed
using the 10D notation \cite{DiVecchia:2011mf}.
Combining it with the
fundamental hypermultiplet part of the action, one gets
\bea
\cL &=&  \Tr\Big(-\fr14 F_{MN}F^{MN}
-\fr{i}2 \lb \G^MD_M \l \Big)
                   -\fr13 if\,\gym\e_{pqr}\f^p\f^q\f^r \nn\\
 &&+ \e^{\dal\dbe}(D_{\muh}\wb_{\dal}) (D^{\muh}w_{\dbe})
       - (\wb_{\dal})\sum_{c=1}^3 (\t^c)^{\dal}{}_{\dbe}
              (\cD_{c}) (w^{\dbe})\nn\\
&& -i\mub \G^{\muh} (D_{\muh} \m)
   +2i\, \Big[\mub(\chi^{\dal}) \e_{\dal \dbe} (w^{\dbe})^b{}_u
       + (\wb_{\dal}) \e^{{\dal}\dbe}(\chib_{\dbe})\, \m \Big]
 \label{10Dsymplus}
\eea
The relationship between gauge fermionic terms in
\rf{sym3q} and $\l$-terms in \rf{10Dsymplus} can be found
in \cite{DiVecchia:2011mf}.

As we have noted in the previous section, the supersymmetry
transformations of the gauge multiplet fields do not involve
any hypermultiplet fields. This implies that the gauge multiplet and the hypermultiplet fields should be treated differently.
Also because of the aforementioned branch matching that implies $w_0=0$, we focus on the gauge multiplet part of the lagrangian, choosing it as $QS_L$.\footnote{One may consider a localization lagrangian for the hypermultiplet, and add it to the total lagrangain as well. We expect that the hypermultiplet part of the contribution to the entropy
should be subleading basically due to supersymmetry.
 }
Let us focus on the adjoint fields,
\bea
{t} QS_L={t} \cL_{adjoint}= {t}\Tr\Big(-\fr14 F_{MN}F^{MN}
-\fr{i}2 \lb \G^MD_M \l \Big)
-\fr{i}3 tf\,\gym\,\e_{pqr}\f^p\f^q\f^r
 \label{localL}
\eea
where $t$ is a localization parameter.

Using this orientation, let us now work with \rf{sym4}.
With the choice of the branch $w_0=0$, the bosonic part of the potential
is given by \rf{sym4} which we quote here,
\bea
V &=& -\Tr\Big(
    \fr14[\f_{\muh},\f_{\nuh}]^2+ \fr12\sum_{c=1}^3\cD_{c}^2
       +\fr12[\f_{\muh},\f_m]^2
+\fr{i}2  \sum_{c=1}^3 \cD_{c}\eta_{cmn} [\f^m,\f^n]
   \Big)  +\fr{i}3 f\gym\,\e_{pqr}\f^p\f^q\f^r \label{pot}
\nn\\
\eea
The ${\cD}$-field equation of \rf{symd0fromsym3} is
\bea
2\cD_c+i\eta_{cmn}[\f^m,\f^n]=0
\eea
Substituting this into \rf{pot},
 the potential becomes
\bea
V &=&- \Tr\Big( \fr14 [\f^M,\f^N]^2-\fr{i}3  f\,\gym\; \e_{pqr}
        \f^p\f^q\f^r         \Big)
\eea
where $M,N$ are 10D indices. This potential was analysed by Myers
in \cite{Myers:1999ps}.
Upon substituting the vacuum solution that represents a
non-commutative configuration, the potential yields
\bea
 V=-\fr{ f^4\gym^4}{32} \, N(N^2-1)
\eea
The entropy is obtained by taking a derivative of this
result with respect to the SYM coupling, and it displays the leading
$N^3$ behavior.

 Finally, let us turn to the fluctuation part and expand \rf{localL} around the non-commuting solution just reviewed. In the saddle point method, we keep
only up to (and including) the quadratic terms,
\bea
\cL_{adjoint,2nd} &=& \Tr\Big(-\fr14 F_{MN}F^{MN}
-\fr{i}2 \lb \G^MD_M \l \Big)|_{\f\ra \f+\f_0,\; 2nd}
+\fr{i}3 f\,\gym\,\e_{pqr}\Tr\,\f^p\f^q\f^r|_{\f\ra \f+\f_0,\; 2nd} \nn\\
\label{localLexpanded}
\eea
The path integral that one should evaluate is
\bea
&&\int dA d\f  d\l\; e^{\int -\fr{1}{4}\Tr (F_{0,mn} F_{0,mn})\;
    +\;\fr{i}3 f\,\gym\,\e_{pqr}\Tr\, \f_0^p\f_0^q\f_0^r}
           e^{- \fr14 \Tr\; f_{\m\n} f^{\m\n}}
  e^{\int-\fr{i}2 \lb \G^M(\pa_M-iA_0^M) \l }\nn\\
     &&      \exp\Big[\int- \fr12 \Tr\,\pa_{\m}\f_{m }\pa^{\m}\f^{m }
 + if\,\gym\,\e_{pqr}\Tr\, \f_0^r\f^p\f^q \nn\\
&&-\fr12 \Tr
        [\f_{0,m},\f_{0,n}] [\f^{m},\f^{n}]
 -\fr12 \Tr
        [\f_{0,m},\f_{n}] [\f^{m},\f^{0n}]
-\fr12 \Tr
        [\f_{0,m},\f_{n}] [\f^{0,m},\f^{n}]
\Big]
\label{pic}
\eea
where $f_{\m\n}\eqv \pa_\m A_\n-\pa_\n A_\m$.
Note that the scalar part and the gauge part decouple.
The remaining task is to show that the rest of the integration does
not change the leading $N^3$ behavior of the classical
 contribution.

The leading $N$ behavior should come from the scalar part of the path integral,
\bea
&& \int d\f\,\exp\Big(\fr{1}{\gym^2}\int - \fr12 \Tr\,\pa_{\m}\f_{m }\pa^{\m}\f^{m }
 + if\gym\,\e_{pqr}\Tr\, \f_0^r\f^p\f^q \nn\\
&&-\fr12 \Tr
        [\f_{0m},\f_{0n}] [\f^{m},\f^{n}]
 +\fr12 \Tr
        [\f_{0m},\f_{n}] [\f^{0n},\f^{m}]
-\fr12 \Tr
        [\f_{0m},\f_{n}] [\f^{0m},\f^{n}]
\Big)\nn \\  \label{sp}
=&& \int d\f\,\exp\Big(\fr{1}{\gym^2}\int - \fr12 \,\pa_{\m}(\f_{m })_{s_1t_1}
                     \pa^{\m}(\f^{m })_{t_1s_1}
+  \fr{f^2(N^2-1)}{16}\d_{t_1t_2}(\f_{n})_{t_1s_2}
                 (\f_{n})_{s_2t_2}  + V_{\f^2}\Big)\nn\\
\eea
where we have defined
\bea
V_{\f^2}&\equiv &-\fr12 \Big(
[\f_{0m}, \f_{0n}]_{s_1s_2}[\f_{m s_2t_2}\f_{nt_2s_1}
             -\f_{n s_2t_2}\f_{m t_2s_1}]\Big)
\nn\\
&&+\fr12 \Big(
\f_{0m s_1t_1} \f_{0n s_2t_2}\f_{nt_1s_2}
                 \f_{m t_2s_1}
- \f_{0ms_1t_1} \f_{0nt_2s_1}\f_{nt_1s_2}\f_{ms_2t_2}  \nn\\
&&\quad\;\; - \f_{0mt_1s_2} \f_{0ns_2t_2}\f_{ns_1t_1}
                 \f_{mt_2s_1}
+\f_{0mt_1s_2} \f_{0nt_2s_1}\f_{ns_1t_1}
                 \f_{ms_2t_2}
    \Big)\nn\\
&&- \fr12 \Big(
\f_{0ms_1t_1} \f_{0ms_2t_2}\f_{nt_1s_2}
                 \f_{nt_2s_1}
+\f_{0mt_1s_2} \f_{0mt_2s_1}\f_{ns_1t_1}
                 \f_{ns_2t_2}
    \Big)\nn\\
&&+{if}\,\gym\,\e_{pqr}(\f_0^r)_{s_1t_1}\f^p_{t_1s_2} \f^q_{s_2s_1}
\label{V2nd}
\eea
It should be possible to fully evaluate the integral since the exponent is of quartic
order in $\f$. For our purpose it
is sufficient to evaluate it perturbatively. Let us view
the first line as the propagator terms and the rest as
vertices. The propagator is given by
 \bea
 <\f_{m s_1t_1}(x)\f_{n s_2t_2}(y)>=\gym^2
  \d_{mn}\Big(\d_{s_1t_2}\d_{s_2t_1}-\fr1{N}\d_{s_1t_1}\d_{s_2t_2}\Big)
 \D_{xy,N}
\eea
\bea
 \D_{xy,N}\equiv \int \fr{dq}{2\pi i}\; \fr{e^{iq(x-y)}}{Nq^2-\fr{f^2}{16} N(N^2-1)}
 \label{propa}
\eea
 One factor of $N$ in the denominator of \rf{propa} could be rescaled away
 to be in line with the usual convention;
with the rescaling, the discussion below would still be valid since all the
vertices would get an extra factor of $\fr{1}{N}$ as well.
In the computation below, we only keep the
$\d_{s_1t_2}\d_{s_2t_1}$ piece since we are interested in the
leading $N$ behavior.
The first few Feynman diagrams that need to be evaluated are
given in Fig. 1.

\begin{figure}
\centerline{
\begin{minipage}[b]{7cm}
             \epsfxsize=7cm
              \epsfbox{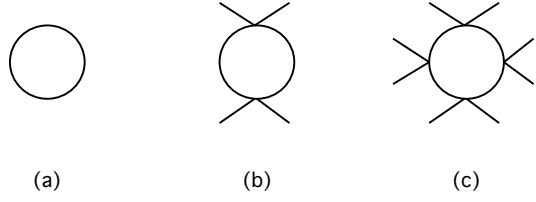}
      \end{minipage}
      }
\caption{one-loop diagrams}
\label{1loopfig}
\end{figure}

For an illustration, let us consider the diagram (b).
We will now show that the leading terms have $N^3$
behavior. Also, it is not difficult to
see from power-counting that the higher order diagrams are at
most of $N$-cubic order.
The diagram (b) corresponds to two insertions of
$V_{\f^2}$. When squared, the leading behavior comes
from the terms in the second bracket in \rf{V2nd}; they yield
\bea
\fr{f^4}{2048}(2N^2+N)(N^2-1)^2\int dxdy\, \D_{xy,N}^2
\eea
The integral yields $N^{-\fr92}$ making the overall leading
power $N^{\fr32}$.
The leading $N$ behavior of the diagrams with more insertions
of $V_{\f^2}$ can be deduced as follows. With more insertions,
the number of the propagators increases as well. At the final
stage of the evaluation, one would compute
\bea
 (N\!\!-\mbox{factors from vertices})\int dp \fr{1}{\Big[Np^2-\fr{f^2}{16} N(N^2-1)\Big]^{\#}}
\eea
The increased power of $N$ in the numerator due to
the presence of more insertions is cancelled by the increased power due
to more propagators in the denominator. For example, the diagram (c) with four insertions
gets $N^{12}$ coming from the vertex contractions and
$N^{-\fr{21}{2}}$ coming from the propagators preserving
the overall $N^{\fr32}$ power. Based on this, we
conclude that the leading terms in the one-loop
contribution have $N^3$ behavior that comes from the classical
contribution.

\subsection{Comments on entropy of a D(-1)/D3 system}

As widely known, the entropy of the extremal D3 brane
configuration has $N^2$ behavior. Let us ponder
the differences between a D0/D4 and a D(-1)/D3 systems.\footnote{
A study of a D(-1)/D3 system can be found, e.g., in \cite{Ito:2007hy}.
}
Although the Myers' term in the D0 brane context has been long-known,
the direct string theoretic realization in the context of general
Dp branes is very recent; it is the eq.(31) in \cite{Hatefi:2012ve}\footnote{
Additional Myers' terms have also been found in \cite{HP}.
} that is based
on the earlier works
\cite{Garousi:2007fk}\cite{Garousi:2008ge}\cite{Hatefi:2010ik}\cite{Hatefi:2012wj}.
We quote it for convenience; in the present notation, it is given by
 \bea
\int d^{p+1}x\; \e^{\m_0\cdots \m_p}\, \tr (F_{\m_0\m_1}\f^k) \pa_k C^{(p-1)}_{\m_2\cdots \m_p}
 \eea
up to a multiplicative constant. The $\m$-index denotes the worldvolume directions and
 the $k$-index, the transverse directions. The Myers' term in the D0 context arises upon
reducing this term to time dimension. There is a
peculiarity in the equation above; when the dielectric effect is realized in string theory context, it could also be realized
through the closed string coupling to
{\em lower} dimensional objects than the dimension of the brane that one
has started with. We will have more on this in the conclusion.
Another implication of the result above is associated with how the effect of D4 branes should be incorporated
in the D0-based description; one should start with D4 branes, and reduce them to
get the D0 dielectric effect. This is, therefore, where an effect of the presence
of D4 branes can be seen in the D0-based description: the presence of the D4 branes -which would not have any effect
in the computation of the partition function otherwise - affects the system through the  Myers terms given above.

For the Myers' terms of a D0/D4 system in the D0-based description,
one should take $p=4$ and reduce this to time dimension.
For a D(-1)/D3 system in the matrix theory description, $p=3$ should
be chosen, followed by reduction to
zero dimension.
The factor $\tr (F_{\m_0 \m_1}\f^k)$ becomes $\tr ([\f_{\m_0},\f_{\m_1}]\f^k)$ which is
totally antisymmetric in $(\m_0,\m_1,k)$. Unlike the $p=4$ case,
$\m_0$ or $\m_1$ must take the time direction in the case of $p=3$.
The whole term would be removed by taking a temporal gauge.
This indicates that, for D(-1)/D3, it might not be the dielectric effect that is
responsible for the $N^2$ behavior.\footnote{ However,
it could be {\em higher $\a'$-order} dielectric effect that is
responsible for the $N^2$ behavior. }

\section{Conclusion}

In this work, we have reproduced the leading $N^3$
entropy behavior from D0 quantum mechanics with the Myers'
terms. The effect of D4 brane is incorporated through the
hypermultiplet in the fundamental representation.
The leading $N^3$ behavior came from the classical
contribution with the rest of the contributions yielding
subleading expressions.

In the D0-based description,
 the presence of the D4 branes gets to introduce additional branches via
 $w$. However, the supergravity $N^3$ behavior corresponds to the branch where $w_0=0$. Therefore, the presence of D4 branes has no effect on the
classical part of the free energy. It does not contribute to the
one-loop part of the partition function either (as seen in the
previous section) by choosing
the localization term solely in terms of the gauge multiplet.
 This does not mean that the $w$-field has no effect on the free
energy in general. It just means that the $w$-field is irrelevant
for the particular class of the supergravity solutions under consideration.

Although natural, there are several assumptions that were made
to derive the results in this paper. Firstly, we used a D0-based description in
the spirit of \cite{Dorey:2002ik};
it is not entirely clear whether the D0-based description would be capable
of capturing the full physics of a D0/D4 system although it would certainly
capture some aspects of it. This matter would be worth looking into.
Secondly, we have noticed in section 4 that the Myers' terms can be realized in a peculiar
manner from the coupling between open and closed strings. It comes from the
closed string coupling to a lower dimensional branes as well that "lie inside" of
the branes that one started with. The lower dimensional branes
can be viewed as soliton solutions of the branes that one started
with. (This seems consistent with the view that the closed string
is taken as a composite state of the open string theory.) A better understanding
of this phenomenon would be desired. The finite temperature effect is another aspect that requires attention.
Finally, it would be interesting to investigate whether the $N^2$ behavior
of extremal D3 branes could be related to a higher order {dielectric} effect.\footnote{Note \cite{Leigh:2010va} in this respect, where the $N^2$ and $N^3$ entropy growth of effective potentials was observed within effective subdeterminant models of $N$ coincident D3 and M2 branes.}
We hope to report on some of these issues in the future.

\vspace{.5in}

\ni {\bf Acknowledgements}\\
EH would like to thank K.S. Narain, F. Quevedo, L. Alvarez-Gaume,
N. Arkani-Hamed, G. Veneziano and N. Lambert. He is especially grateful to
Rob Myers for very beneficial correspondences.
{Work of AJN is supported in part by the Joint DFFD-RFBR Grant \#F40.2/040.} He thanks Vladimir Ivashuk for discussion and Anastasios Petkou for communication.
IP thanks Jean-Emile Bourgine, Goro Ishiki, Satoshi Nawata, Satoshi Watamura,
Satoshi Yamaguchi and Hynn-Seok Yang for their discussions/correspondences.


\end{document}